\documentclass[prl,aps,twocolumn,showpacs,amsmath,amssymb]{revtex4}

\usepackage{graphicx}
\usepackage{color}

\newcommand{\be}{\begin{equation}}
\newcommand{\ee}{\end{equation}}

\renewcommand{\vec}[1]{\mathbf{#1}}

\begin{document}

\title{Confining stationary light: Dirac dynamics and Klein tunneling}
\author{ J. Otterbach$^1$,  R. G. Unanyan$^{1,2}$, and M. Fleischhauer$^1$}
\affiliation{$^{1}$Fachbereich Physik, Technische Universit\"{a}t Kaiserslautern, 67663,
Kaiserslautern, Germany}
\affiliation{$^{2}$Institute for Physical Research Armenian National Academy of Sciences,
Ashtarak-2 378410, Armenia}
\begin{abstract}
We discuss the properties of 1D stationary pulses of light in atomic ensemble
with electromagnetically induced transparency in the limit of tight spatial confinement.
When the size of the wavepacket becomes comparable or smaller
than the absorption length of the medium, it must be described by a two-component vector which obeys the
one-dimensional two-component Dirac equation with an effective mass $m^*$ and effective speed
of light $c^*$.  Then a fundamental lower limit
to the spatial width in an external potential arises from Klein tunneling and is
given by the effective Compton length $\lambda_C = \hbar/(m^* c^*)$.
Since $c^*$ and $m^*$ can be externally controlled and can be made
small it is possible to observe effects of the relativistic dispersion
for rather low energies or correspondingly on macroscopic length scales.
\end{abstract}


\maketitle


When photons are confined to a volume smaller than the wavelength cubed
their interaction with atoms is dominated by quantum effects.
This principle has been exploited in cavity quantum electrodynamics, where the light is confined
by means of low-loss micro-resonators \cite{cavity-QED}. The tight confinement results in a strong
coupling which can be used e.g. to build quantum gates between photonic qubits.
Yet with a decreasing resonator volume it becomes more and more difficult to maintain high $Q$ values.
However, as shown by Bajcsy et al.  \cite{Bajcsy-Nature-2003,Andre}, it is possible to create spatially confined
quasi-stationary pulses of light with very low losses without the need of a resonator by means of electromagnetically
induced transparency (EIT) \cite{Harris-Physics-Today-1997,Fleischhauer-RMP-2005} with
counterpropagating control fields. For a weak confinement in the
longitudinal direction, stationary light is well described by a Schr\"odinger-type equation with complex mass for a
normal mode of the system, the stationary dark-state polariton \cite{Zimmer-OptComm-2006,Zimmer-PRA-2008,Chong-PRA-2008}.
We here
show that this is no longer the case for stronger spatial confinement, i.e. when the characteristic length of the
photon wavepackets becomes comparable or smaller than the absorption length of the medium. Here a description in terms of
a two-component vector obeying a one-dimensional Dirac equation becomes necessary. The two characteristic parameter
of this equation, the effective mass $m^*$, and the effective speed of light $c^*$ depend on the strength of the
EIT control fields and can be made many orders of magnitude smaller than the vacuum speed of light $c$ and respectively
the mass of the atoms forming the EIT medium. As a consequence effects of the relativistic dispersion can arise already
at rather low energy scales. On one hand this leads to a fundamental lower limit for the spatial confinement of
stationary light in an external potential given by the effective Compton length $\lambda_C= \hbar/(m^* c^*)$ which
due to the smallness of $m^* c^*$ can become large.
On the other hand it opens the possibility to study relativistic effects such as Klein tunneling \cite{Klein} and Zitterbewegung
which regained a lot of interest recently in connection with electronic properties of graphene \cite{graphene} and ultra-cold
atoms in light- or rotation-induced gauge potentials \cite{atoms}.

\begin{figure}[hbt]
    \begin{center}
   \includegraphics[width=5 cm]{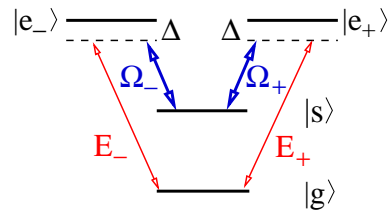}
      \caption{(color online) stationary light scheme: The interaction of double $\Lambda$ atoms driven by
two counter-propagating control fields of (equal) Rabi frequency $\Omega$
and opposite circular polarization with two counter-propagating probe fields $E_\pm$ of corresponding polarization
generate a quasi-stationary pattern of the probe fields.
}
      \label{fig:system}
    \end{center}
\end{figure}

We here consider an ensemble of quantum oscillators with a double-$\Lambda$ structure of
dipole transitions as shown in Fig.\ref{fig:system}.
The ground state $|g\rangle$ and the meta-stable state $|s\rangle$ are coupled via
Raman transitions through the excited states $|e_\pm\rangle$ by two counterpropagating control
laser of opposite circular polarization and Rabi-frequencies $\Omega_\pm$ and two counterpropagating
probe fields $E_\pm$ again  of opposite circular polarization.
Both $\Lambda$ schemes are assumed to be in two-photon resonance with the ground state transition, which guarantees EIT.
Furthermore the control fields are taken homogeneous, constant in time and of equal strength
$\Omega_+ =\Omega_+^*=\Omega_-=\Omega_-^*$. As shown in \cite{Zimmer-PRA-2008}
the control fields generate a quasi-stationary pulse of light. For the present
discussion we restrict ourselves to a one-dimensional dynamical model.
One can show that the transverse dynamics occurs on a much longer time
scale than the longitudinal one, which justifies this simplification
\cite{Otterbach-thesis}.

We introduce normalized field amplitudes that vary slowly in space and time , $E_\pm(\vec r,t)
=\sqrt{\frac{\hbar\omega}{2\varepsilon_0}} \Bigl({\cal E}_\pm(\vec r,t) \exp\{-i(\omega t \mp kz)\} + h.a.\Bigr)$,
and continuous atomic-flip operators $\hat\sigma_{\mu\nu}(\vec r,t) = \frac{1}{\Delta N}\sum_{j\in \Delta V(\vec r)} \hat \sigma^j_{\mu\nu}$, with $\hat\sigma_{\mu\nu}^j
\equiv |\mu\rangle_{jj}\langle \nu|$ being the flip operator of the $j$th atom.
The sum is over all $\Delta N$ atoms in a small volume $\Delta V$ around position $\vec r$.
Then the dynamical equations read in the linear response limit, i.e. for a low probe light intensity
\begin{eqnarray}
\frac{\partial}{\partial t}\hat \sigma_{gs} &=& i\Omega_+ \hat\sigma_{ge_+} +i\Omega_-\hat\sigma_{ge_-},
\label{eq:sigma-gs}\\
\frac{\partial}{\partial t}\hat\sigma_{ge_\pm} &=& -(\gamma+i \Delta)\hat\sigma_{ge_\pm}+i\Omega_\pm\hat\sigma_{gs}
+i g\sqrt{n}\, {\cal E}_\pm,\label{eq:sigma-ge}\\
ig\sqrt{n}\hat \sigma_{ge_\pm} &=& \left[\frac{\partial}{\partial t}\pm c\frac{\partial}{\partial z}\right] {\cal E}_\pm.
\end{eqnarray}
Here we have introduced the common single-photon detuning of the upper states from the control
and probe field transitions $\Delta$.
$n$ is the atom density and $g=\frac{\wp}{\hbar}\sqrt{\frac{\hbar \omega}
{2\varepsilon_0}}$ the common coupling constant of both probe fields with $\wp$ denoting the
respective dipole matrix element. $\gamma$ is the transverse decay rate of the optical dipole transitions $|e_\pm\rangle - |g\rangle$ and we have used that in the linear response limit $\hat\sigma_{gg}\approx 1$.

Adiabatically eliminating the atomic variables from the equations of motion leads to the
shortened wave equations for the forward $({\cal E_+})$ and backward $({\cal E_-})$ propagating
field components
\begin{eqnarray}
\left[\frac{\partial}{\partial t}\pm c\frac{\partial}{\partial z}\right]
{\cal E}_\pm &=& - \frac{g^2 n}{\Gamma} {\cal E}_\pm + \frac{g^2 n}{2\Gamma}\left({\cal E}_++{\cal E}_-\right)\nonumber\\
&& -\tan^2\theta \left(\frac{\partial}{\partial t}{\cal E}_+ +\frac{\partial}{\partial t}{\cal E}_-\right),
\end{eqnarray}
where $\Gamma = \gamma +i\Delta$, and $\tan^2\theta = g^2 n/\Omega^2$, with $\Omega^2 = \Omega_+^2+\Omega_-^2$.
If the characteristic length scale of the probe fields is large compared to the absorption length of the medium,
$L_{\rm abs} = \gamma c / (g^2 n)$, it is convenient to introduce sum and difference normal modes ${\cal E}_S =
({\cal E}_++{\cal E}_-)/\sqrt{2}$ and ${\cal E}_D =({\cal E}_+-{\cal E}_-)/\sqrt{2}$. Expressing the equations of motion in terms of these normal modes and subsequently adiabatically eliminating the fast decaying difference normal mode ${\cal E}_D$ leads to a Schr\"odinger-type
equation with a complex effective mass \cite{Zimmer-OptComm-2006,Zimmer-PRA-2008}.
\begin{eqnarray}
i\hbar \frac{\partial}{\partial t}\, {\cal E}_S = - \frac{\hbar^2}{2 m^*} \left(1+i\frac{\gamma}{\Delta}\right)\frac{\partial^2}{\partial z^2}\, {\cal E}_S.
\label{eq:Schroedinger}
\end{eqnarray}
Here $m^*$ denotes the real part of the effective mass
\begin{equation}
m^* = \frac{\hbar}{2 L_{\rm abs}}\frac{1}{c \cos^2\theta} \frac{\gamma}{\Delta}
= m \frac{v_{\rm rec}}{v_{\rm gr}} \frac{\gamma}{\Delta} \frac{1}{2 k L_{\rm abs}}.
\end{equation}
As comparative scales we have introduced in the second equation the mass $m$ and the recoil
velocity $v_{\rm rec} = \hbar k /m$ of an atom and the group velocity of EIT $v_{\rm gr}= c\cos^2\theta$.

On the other hand if the characteristic length scale of the probe fields becomes comparable to the
absorption length non-adiabatic couplings between the sum and difference mode are relevant and the
elimination of the difference mode is no longer valid.
Instead one has to keep both amplitudes which can be collected in
a two component vector $\underline{\widetilde{\cal E}} = ({\cal E}_+,{\cal E}_-)^\top$. The equation of motion
can then be written in the compact form
\begin{eqnarray}
\left({\sf A}\, \frac{\partial}{\partial t} +{\sf B}\,
\frac{\partial}{\partial z}\right)\underline{\widetilde{\cal E}} = {\sf C}\, \underline{\widetilde{\cal E}}
\end{eqnarray}
where ${\sf A} = \mathbf{1} + \frac{1}{2}\tan^2\theta (\mathbf{1}+\, \sigma_x)$, ${\sf B}
= c \, \sigma_z$, and ${\sf C}= (\sigma_x -\mathbf{1})g^2 n/(2\Gamma) $, $\sigma_x$ and $\sigma_z$ being the
Pauli matrices.
Applying the transformation
\begin{equation}
\underline{\cal E} = \exp\left\{\beta\, \sigma_x\right\}\, \underline{\widetilde {\cal E}},
\end{equation}
with $\tan(2\beta) = (1-\cos^2\theta)/(1 +\cos^2\theta)$ finally yields for large single-photon
detuning $|\Delta|\gg \gamma$, i.e. for a negligible imaginary part of the effective mass:
\begin{eqnarray}
i \hbar \frac{\partial}{\partial t}\underline{\cal E} = \left(- i\hbar \, c^*\, \sigma_z\,
\frac{\partial}{\partial z} + m^* c^{*2}\, \sigma_x\right)\underline{\cal E}.
\label{eq:Dirac}
\end{eqnarray}
Here we have removed an irrelevant constant term by a gauge transformation.
Eq.(\ref{eq:Dirac}) has the form of a two-component, one-dimensional, massive
Dirac equation. The effective speed of light $c^*$ in eq.(\ref{eq:Dirac})
\begin{equation}
c^* = c\, \cos\theta = \sqrt{v_{\rm gr} c}
\end{equation}
which in EIT media can be varied over a large range and can be much smaller than the vacuum
speed of light.
It should be noted that despite the formal equivalence of equation (\ref{eq:Dirac})
to the two component Dirac equation the fundamental quasi-particle excitations of the
light matter interaction are bosons \cite{Zimmer-PRA-2008} and can e.g. undergo Bose
condensation \cite{Fleischhauer-PRL-2008b}. Equation (\ref{eq:Schroedinger}) is of course recovered from
eq.(\ref{eq:Dirac}) in the low energy limit of long wavelength excitations.

The characteristic
length scale at which relativistic effects become important is the Compton length
\begin{equation}
\lambda_C \equiv \frac{\hbar}{m^* c^*} = 2 L_{\rm abs} \, \frac{\Delta}{\gamma} {\cos\theta}.
\end{equation}
While for electrons $\lambda_C$ is on the order of pico-meters, for stationary light it can become
rather large. It can exceed the absorption length if the EIT group velocity is sufficiently large, i.e.
$v_{\rm gr}/c > \gamma/\Delta$. Since typical values  for the optical depth $OD=L/L_{\rm abs}$ of
EIT media are in the range between a few and a few hundred, $\lambda_C$ can be a sizable
fraction of the medium length $L$ and thus can become macroscopic.

\begin{figure}[hbt]
    \begin{center}
   \includegraphics[width=5.5 cm]{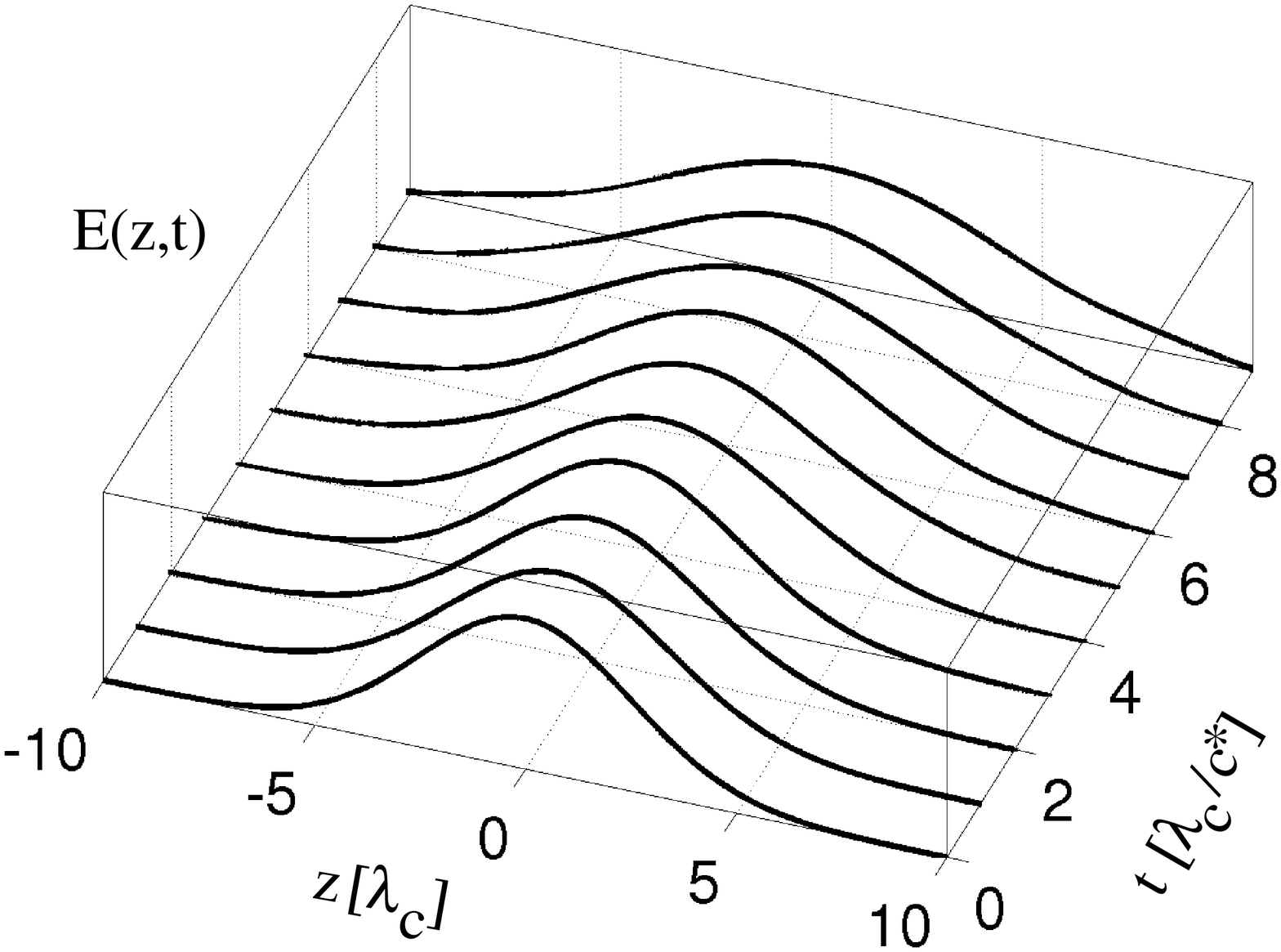}
\vspace*{.3cm}
   \includegraphics[width=5.5 cm]{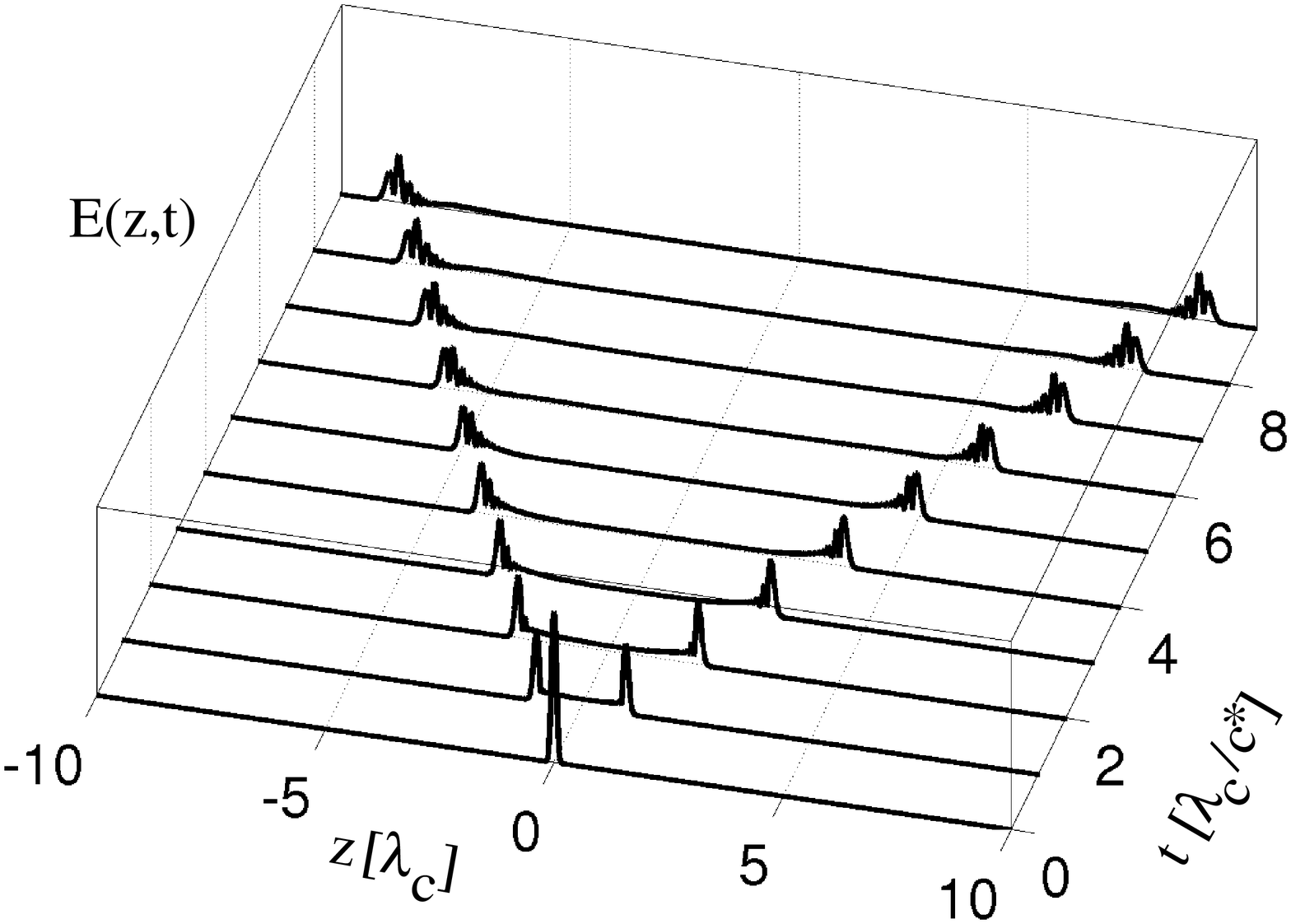}
      \caption{{\it top:} diffusive expansion of an initial gaussian stationary light pulse with
width $L_p=2.5 \lambda_C$. {\it bottom:} the same for an initial width of $L_p=0.05 \lambda_C$.
Results are obtained from numerical solutions of the full Maxwell-Bloch equations
which agree very well with solutions of the Schr\"odinger equation (\ref{eq:Schroedinger}) (top plot)
and the Dirac equation (\ref{eq:Dirac}) (bottom plot). The parameters are
${\gamma}/{\Delta}=0.01$, ${\Omega_\pm}/{\Delta}=0.2$, $\cos\theta=0.9975$ and hence ${\lambda_C}/{L_{\rm abs}}=199.5$
}
      \label{fig:dynamics}
    \end{center}
\end{figure}

The free evolution of a quasi-stationary pulse of light is quite different in the two limits
$L \gg \lambda_C$, eq.(\ref{eq:Schroedinger}), and $L\lesssim \lambda_C$,
eq.(\ref{eq:Dirac}), which allows for a simple experimental distinction of the two regimes.
This is illustrated in Fig.\ref{fig:dynamics} which shows the dynamics of a wavepacket
obtained from a numerical solution of the full Maxwell-Bloch equations, which agrees with
the dynamics from the effective Schr\"odinger and Dirac equations.
One recognizes in the first case the familiar
slow dispersive spreading, while in the second case two split wavepackets
emerge which move outward with the effective speed of light $c^*$.

\begin{figure}[hbt]
    \begin{center}
   \includegraphics[width=5.5 cm]{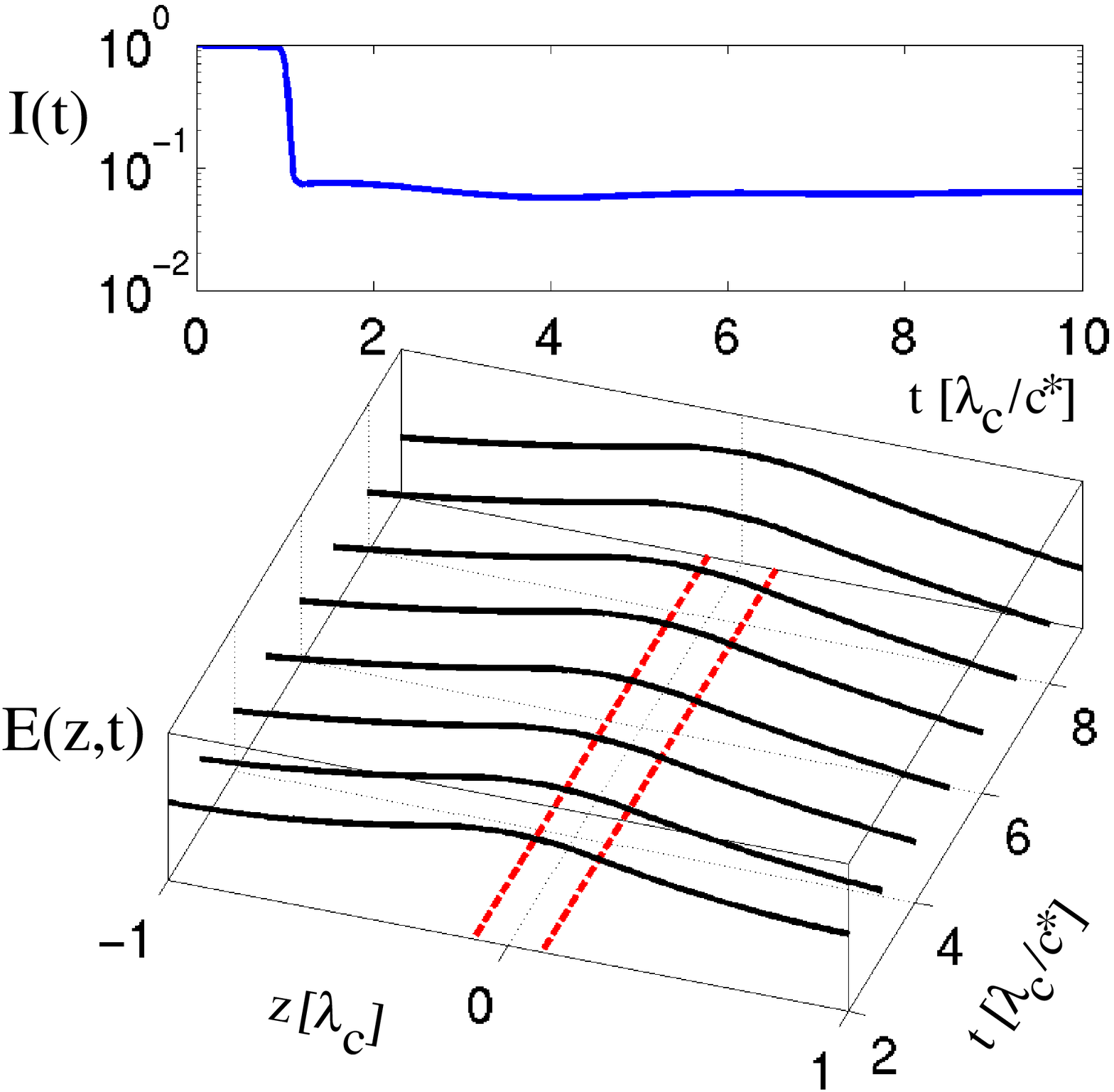}
\vspace*{.3cm}
 \includegraphics[width=5.5 cm]{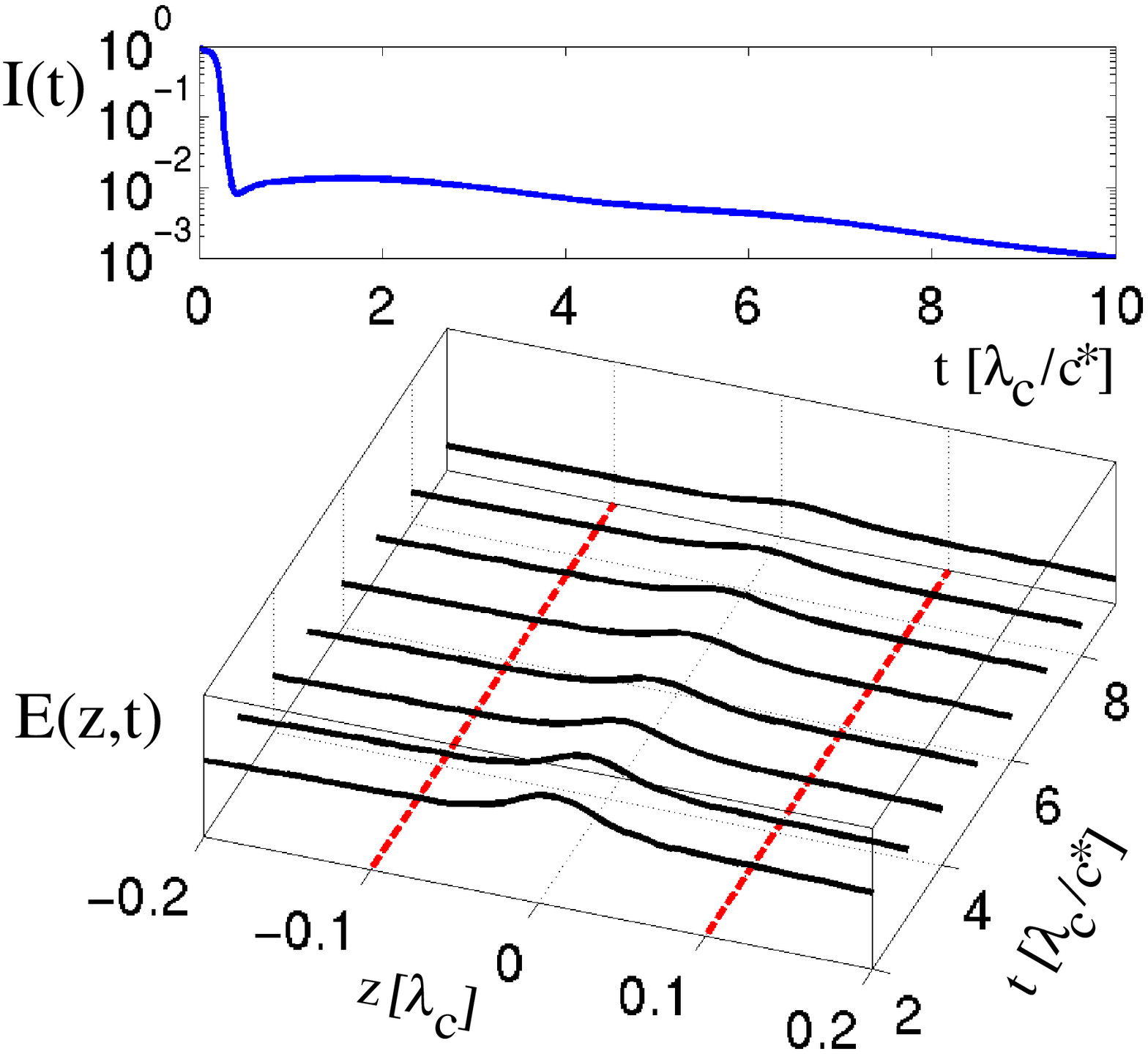}
      \caption{(Color online) evolution of an initial gaussian
stationary light pulse with initial width $L_p=0.05\lambda_c$ in a square well potential (the extension of which is indicated by dashed red lines) with $a=0.1\lambda_C$ and depth $U_0=1.875 m^*c^{*2}$ ({\it top}), and $U_0=3.125m^*c^{*2}$ ({\it bottom}) respectively. Besides initial losses due to mismatch of the wavefunctions (see insets at the top) a fast decay due to Klein tunneling is apparent in the second case.
}
      \label{fig:confinement}
    \end{center}
\end{figure}

It is well known that a relativistic wave equation does not permit a confinement
of a wavepacket to less than the Compton length. E.g.
a square-well potential
\begin{equation}
U(z) = \left\{
\begin{array}{cc}
 -U_0 &\quad |z|\le a\\ 0 &\quad |z| > a
\end{array}\right.
\end{equation}
with $a\to 0$ and $U_0\to \infty$ such that $U_0 a =$ const.
has lowest energy eigensolutions with energy
\begin{eqnarray}
E = \pm m^* c^{*2} \, \cos\left(\frac{U_0}{m^* c^{*2}}\frac{a}{\lambda_C}\right).
\end{eqnarray}
The corresponding eigensolutions
have the form
\begin{equation}
{\cal E} = {\cal E}^{(\pm)} \exp\left[- \frac{m^* c^* |z|}{\hbar}\left\vert
\sin\left(\frac{U_0}{m^* c^{* 2}} \frac{a}{\lambda_c}\right)\right\vert\right]
\end{equation}
Thus the characteristic confinement length $L_{\rm conf}$ reads
\begin{equation}
L_{\rm conf} = \lambda_C \, \frac{1}
{\left\vert\sin\left(\frac{U_0}{m^* c^{* 2}} \frac{a}{\lambda_C}\right)\right\vert} \ge \lambda_C
\end{equation}
which is always larger than $\lambda_C$. One can show in general that any
eigensolution of any ("electrostatic") confining potential has a minimum size of $\frac{\lambda_{C}}{2}$. If solutions with
an energy exceeding $\mp m^* c^{*2}$ exist they are resonances which have a finite width due to Klein tunneling
into the negative (positive) energy continuum \cite{Klein}.
Depending on the form of the confining potential the corresponding decay
rate can however be small.
The effect of Klein tunneling is illustrated for confined stationary light in Fig.\ref{fig:confinement}.
Here an initial gaussian stationary wavepacket of initial width $L_p=0.05\lambda_c$ is considered
in a potential well with $a= 0.1\lambda_c $  and potential depth $U_0=1.875 m^*c^{*2}$ (top) and $U_0=3.125m^*c^{*2}$ (bottom). Due to the mismatch of the initial wavepacket with the bound-states of the potential there is
some initial loss. After some time the pulse shape remains rather
unchanged however in the first case while it displays continuing decay in the second
due to Klein tunneling. The plots are obtained from a solution of the full 1D Maxwell-Bloch equations with an additional potential generated by a finite, space-dependent two-photon detuning which show very good agreement with the solutions of the
corresponding Dirac equation.

There is another well-known effect of the Dirac dynamics which can be observed in
the present system. If we consider an initial stationary gaussian wavepacket and switch the
relative phase between the two counter-propagating drive fields instantaneously from
$0$ to $\pi/2$, the relative motion of the two emerging wavepackets
 is superimposed
by an oscillation, known as Zitterbewegung.
After the $\pi/2$ phase flip the
initial wavepacket reads in $k$ space
\begin{eqnarray}
\underline{\widetilde {\cal E}}(k,t=0) = \frac{1}{\sqrt{\sigma_k} \pi^{1/4}} \exp\left\{-\frac{k^2}{2\sigma_k^2}\right\}
\left(\begin{array}{c}
1 \\ i
\end{array}
\right)
\end{eqnarray}
Then in the large-time limit $t\gg \hbar/(m^* c^{*2})$ one finds for the center of mass of the two wavepackets
(setting $\gamma/\Delta =0$)
\begin{eqnarray}
\langle z(t) \rangle &=& \frac{\sqrt{\pi}}{\sigma_k \cosh(2\beta)} \times
\label{eq:center}
\\
&&\times\left[ 1-\left(\pi \frac{m^* c^{*2}}{\hbar} t\right)^{-1/2}
\cos\left(\frac{2 m^*c^{*2}}{\hbar} t +\frac{\pi}{4}\right)\right]\nonumber
\end{eqnarray}
which shows the characteristic oscillation with frequency $2 m^* c^{*2}/\hbar$.
\begin{figure}[!t]
    \begin{center}
   \includegraphics[width=.48\textwidth]{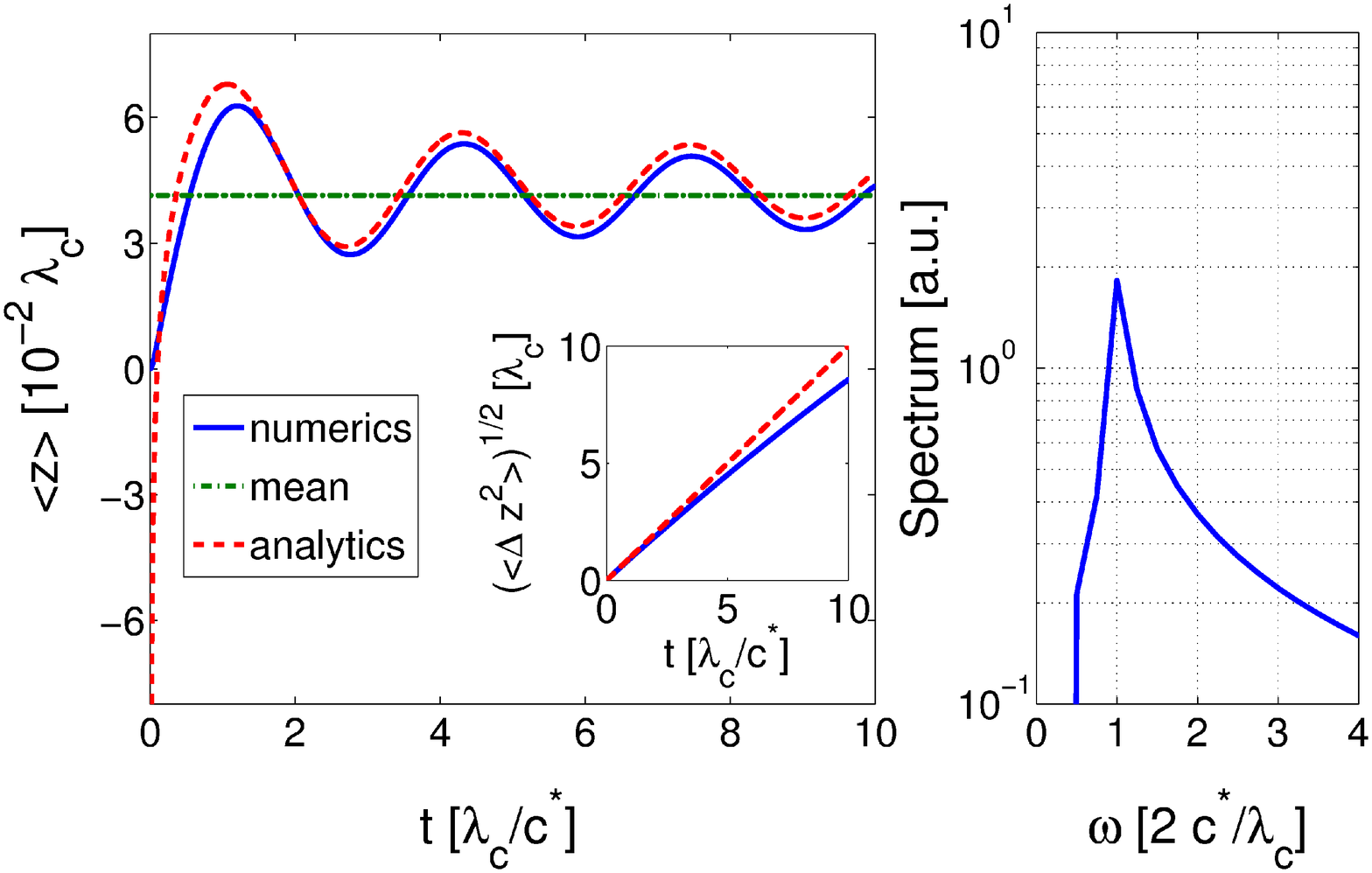}
      \caption{(Color online) {\it left:} center of mass of emerging left and right
moving wavepackets obtaind from solution of Maxwell-Bloch equations
of initial gaussian wavepacket of width $\Delta z = 10 L_{\rm abs} \ll  \lambda_C = 199.98 L_{\rm abs}$
(solid blue line) and from solution (\ref{eq:center}) of the
Dirac equation (dashed red line).
 At $t=0$ a flip of the relative phase
of the drive fields was assumed corresponding to ${\cal E}_+(0) = i {\cal E}_-(0)$. Parameters are as in fig.\ref{fig:dynamics}
{\it right:} Fourier-spectrum of $\langle z(t)\rangle$
showing the pronounced peak at $2 m^* c^{*2}/\hbar$.
}
      \label{fig:Zitter}
    \end{center}
\end{figure}
Fig.\ref{fig:Zitter} shows the center of mass of a pair of left and right moving wavepackets for an initial gaussian pulse obtained from a numerical solution of the 1D Maxwell-Bloch equations as well as the analytic result (\ref{eq:center}) obtained from the Dirac equation. The Zitterbewegung can be observed e.g. by detecting the overlap of the left and right moving wavepackets after exiting the medium. A typical scale of the amplitude of the Zitterbewegung would be ${L_{\text{abs}}}/{10}\sim0.1$cm which should be much easier to observe than the values in the range from 1-100nm predicted for cold atoms \cite{atoms}, graphene \cite{graph}, ions \cite{ions} and photonic crystals \cite{Photonic}.

In summary we have shown that stationary light in the limit of tight longitudinal spatial confinement must be described by a two-component, one-dimensional Dirac equation with effective mass and effective speed of light that can be controlled externally and that can be much smaller than the corresponding values for atoms and light in vacuum. As a consequence relativistic effects related to the Dirac dispersion can be observed at rather low energy scales or respectively at rather large length scales. One immediate consequence of the latter is the impossibility to spatially compress a stationary light pulse below the Compton length. Moreover in contrast e.g. to electrons in graphene, interactions between stationary-light polaritons
are very week. Thus stationary light may be employed to observe relativistic phenomena related to the Dirac dynamics in the
absence of interactions under experimentally realistic conditions.

\vfill

\end{document}